\documentclass[10pt,a4paper]{article}
\usepackage{mathrsfs}
\usepackage{fancyhdr}
\usepackage[top=30truemm,bottom=30truemm,left=20truemm,right=50truemm]{geometry}
\usepackage{indentfirst}
\usepackage{authblk}
\usepackage{graphicx}
\usepackage{bm,color}

\renewenvironment{abstract}{\begin{quotation}{\normalsize }}{\end{quotation}}
\makeatletter
\newcommand{\email}[1]{\newcommand{\@email}{E-mail: #1}}
\renewcommand{\maketitle}{
\newpage\null
    \vspace{2em}
 {\LARGE\bfseries\noindent\ignorespaces\@title\par}
 \vspace{1em}%
 {\large\noindent\ignorespaces\@author\par}
 \vspace{2mm}
 {\normalsize\noindent\ignorespaces\@email\par}
\vspace{1em}
}

\renewcommand\@author{\ifx\AB@affillist\AB@empty\AB@author\else
      \ifnum\value{affil}>\value{Maxaffil}\def\rlap##1{##1}%
     \vspace{1mm} \AB@authlist\\\AB@affillist
    \else  \AB@authors\fi\fi}
\makeatother
\usepackage{amsmath,amssymb}
\title{Nuclear Reactions and Superfluid Time Dependent Density Functional Theory}
\author[1,2]{Piotr Magierski}
\affil[1]{Faculty of Physics, Warsaw University of Technology,
ulica Koszykowa 75, 00-662 Warsaw, POLAND}
\affil[2]{Department of Physics, University of Washington, Seattle,
WA 98195--1560, USA}
\email{piotrm@uw.edu}
\begin{document}
\maketitle

\begin{abstract}
{\bf Abstract: The extension of Time Dependent Density Functional Theory (TDDFT) to superfluid systems is discussed in the context of nuclear reactions
and large amplitude collective motion. }

\end{abstract}

{\bf keywords:}  density functional theory, pairing, nuclear reactions, Coulomb excitation.

\section*{Time Dependent Density Functional Theory - general remarks}

Density Functional Theory (DFT) has become nowadays a standard theoretical tool for studies of interacting 
many-body Fermi systems\cite{Kohn:1999,Dreizler:1990lr,Eschrig:1996,Parr:1989}. 
It offers a universal and formally exact approach, which had enormous practical successes. In the field of condensed matter 
it is widely used whenever properties of electronic systems need to be determined \cite{Picket:1989,Brack:1993,Brivio:1999,Freysoldt:2014}. 
There is however a significant difference  between DFT
and other theoretical tools of quantum many-body physics. The latter are usually designed in a way which allows to estimate their applicability and 
also provide a method to systematically improve the predictions and obtain greater accuracy. 
On the contrary, the central object in DFT is the energy density functional which 
is merely proved to exist by means of the Hohenberg-Kohn theorem \cite{Kohn:1964}.
The theorem states that certain unique energy density functional provides the energy of the ground state and spatial density distribution 
as a result of some minimum principle. But it neither offers any method of construction of such functional eg. starting from interparticle interactions,
nor it guaranties that the functional can be written in an analytic form, which is crucial for any practical calculations. 
In electronic systems the situation is nevertheless much simpler as the interelectronic interaction is well known.
Consequently, guided by Hartree-Fock approximation, one may construct the main component of the functional
and only the exchange and correlation energy contributions need to be specified.
Their form is usually extracted from ab-initio Quantum Monte Carlo (QMC) calculations for uniform systems \cite{Ceperley:1978, Ceperley:1980}.

The case of atomic nuclei is far more complicated. Two types of particles: neutrons and protons, need to be taken into account in the description
of the system. The nuclear interaction is quite complex, involving many terms,  including also the three-body part. Moreover, an atomic nucleus
is a selfbound system, which prevents the straightforward application of the Hohenberg-Kohn theorem.
Therefore it is more difficult to quantitatively justify a given choice of the functional. The nuclear energy density functionals have various forms, 
the most popular being the Skyrme functional (including variety of parametrizations), which despite of known shortcomings is
still widely used (see eg. \cite{Tarpanov:2014,Rodriguez-Guzman:2014,Baldo:2013,Fayans:2001,unedf} and references therein).
Although DFT is based on rigorous theorems and a hierarchy of increasingly accurate approximations can be constructed, such as the
local density approximation (LDA), generalized gradient approximations (GGA) and hybrids of exact exchange with GGA\cite{Becke1988, Becke1988a, Becke1988b, Becke1988c}, there is no general method of calculating corrections  involving the information about interparticle
interaction.  More systematic approach to construction of the energy functional can be found in Refs. \cite{Rai2014, Rai2014a, Rai2014b, Furn2010}.

Nevertheless the simple scheme offered by the energy density functional theory is very atrractive, as in the DFT instead of searching for the wave function of an 
$N-$particle system, which depends on $3N$ variables, one solves a system of $N$ nonlinear, coupled partial differential equations. 
This simplification is achieved by introducing the Kohn-Sham (K-S) scheme,
where the density is expressible through the set of orbitals which are determined from variational principle \cite{KS65}.
Consequently the minimization of the functional leads to set of equations for the orbitals defining the density distribution.
The strict formulation of DFT limits its applicability to the ground-state properties of the system.
However in the context of nuclear reactions the proper treatment of excited states is crucial. It can be achieved
with an extension of the DFT to include time evolution.
Time-dependent density functional theory (TDDFT) is an universal approach to the
quantum many-body dynamics (see \cite{Ullrich:2012,Marques:2012,Oni2002} and references therein). It means that TDDFT can be used to describe
nonstationary situations in systems consisting of nuclei, atoms, molecules, solids, or nanostructures. 
TDDFT applies the same philosophy as DFT to time-dependent problems.  
The  Runge-Gross theorem, which  is the time-dependent counterpart of 
the Hohenberg-Kohn theorem proves that
if two $N$-fermion systems evolve from the same initial state, but are subject to two different time-dependent potentials,
their respective time-dependent densities will be different \cite{Runge:1984mz}.
There is however an important problem which troubled TDDFT over many years and is related to the definition of the exchange potential,
which may exhibit nonlocality in time and in principle violate causality principle.
The so-called causality paradox has been resolved in a series of papers \cite{Rajagopal:1996,Leeuwen:1998,Leeuwen:2001,Mukamel:2005,Vig2008}, 
but it also implies that the exact expression of the exchange potential as a functional of the density is unknown and requires
certain approximations. It is however the only fundamental approximation in TDDFT.

Similarly to static K-S equations one can define time dependent
Kohn-Sham equations that describe non-interacting particles that evolve in a time-dependent
Kohn-Sham potential, and produce the same density as that of the interacting system of interest. Thus, just as in
the ground-state case, the time dependent Schroedinger equation is replaced by much
simpler set of equations to solve.
TDDFT can be, and usually is, used to obtain excited states be means of linear response theory (\cite{Elliott:2009} and references therein). 
However, there are important differences as well, and there are features of TDDFT that are unique to the time-dependent case.
First of all the ground-state DFT is based on the variational minimum principle. In the time-dependent
case, there is no analogous minimum principle. It is possible to derive
the formal framework of TDDFT from a stationary-action principle, but
in contrast to DFT, where the ground-state energy is the quantity of central
importance, the action is practically of no interest in itself \footnote{The uniqueness of the stationary-action point remains unproven.}. 
Another new feature of TDDFT are currents,
which need to be added to the description of evolving system\cite{Xu1985,Leeuwen:1999}. Moreover
the feature that have no counterpart in static DFT is that the time-dependent 
exchange-correlation potential contains memory term\cite{Dob1997, Dob1997a}. Namely, the potential
at time $t$ depends on densities $\rho(r\rq{}, t\rq{})$ at earlier times, where
$t\rq{} \le t$. This memory is, in principle, infinitely long-ranged. Unfortunately very little is known
about the memory term and the most common approximation  is the adiabatic
approximation, which ignores all memory effects.
This is obviously very convenient, as it allows to switch easily from DFT to TDDFT, but
imposes serious limitations on the theory. The effects which are incorrectly described 
as a result of this approximation are collective energy dissipation processes \cite{Ullrich:2012}. 

In the context of nuclear reactions the typical situation in which the TDDFT can be used is the following:
a system is initially in a ground state (obtained within the standard DFT) and is then acted upon 
by a perturbation that drives it out of equilibrium. The external perturbations in the nuclear system
can be of various origins: they can be caused by photon absorption, by neutron capture, or
the perturbation can arise as an interaction between the projectile and the target nucleus, which are
initially in their ground states. It has to be emphasized that the resulting deviations from equilibrium, can be arbitrarily strong. 
TDDFT can be applied both in the linear-response regime
(where it provides information about excitation energies and spectral properties) as well as
in the nonlinear regime, where the external perturbations can be strong enough to
compete with, or even override the internal interactions that provide the structure and
stability of matter. This is of particular interest for the induced nuclear fission processes, which one would like
to describe within TDDFT.

The typical procedure used in the context of TDDFT is the following:
 \begin{itemize}
\item Prepare the initial state, which is usually the ground state (in principle, one
can start from any initial state, but non-ground states or even non-stationary initial
states are rarely considered and more difficult to obtain in practice). 
This can be achieved by solving  static Kohn-Sham equations for a nucleus (or nuclei if
more than one is involved in the reaction process) ,  to get a
set of ground-state Kohn-Sham orbitals and orbital energies.
\item The time evolution can be obtained by applying certain external field simulating
eg. the photon absorption, or through generating nonzero velocities of nuclei towards each other
by performing global phase change of orbitals corresponding to
transformation to a moving inertial frame. 
Then one solves the time-dependent Kohn-Sham equation from the initial time to
the desired final time. The
time propagation of the orbitals gives the time-dependent density.
\item During time evolution  one may calculate the desired observable(s) as 
functionals of $\rho(r, t)$.
\end{itemize}

\section*{Superfluidity and Density Functional Theory}

The existence of superfluidity has been experimentally confirmed in a large number of systems, 
including various condensed matter systems, $^{3}$He and $^{4}$He liquids,
nuclei and neutron stars, and both fermionic and bosonic cold atoms in traps. It is also predicted to show up in dense quark matter. 
In the case of low energy nuclear reactions, in particular when
non-magic medium or heavy nuclei are involved, the proper treatment of superfluidity is crucial and the conventional
DFT descripiton has to be extended.
The first attempt to develop the formal framework of DFT for superconductors has been triggered
by the discovery of high-temperature superconductivity \cite{Oli1988, Wack1994}.  
The extension requires an introduction of an anomalous density 
$\chi({\bf r}\sigma, {\bf r\rq{}}\sigma\rq{})=\langle \hat\psi_{\sigma\rq{}} ({\bf r}\rq{}) \hat\psi_{\sigma}({\bf r}) \rangle$
($\sigma$ denotes the spin degrees of freedom), which play the role of the superconducting order parameter.
The pairing potential is then  formally defined as a functional  derivative of the energy functional with respect to $\chi$:
\begin{equation}
\Delta ({\bf r}\sigma, {\bf r\rq{}}\sigma\rq{})=-\frac{\delta E(\rho, \chi)}{\delta \chi^{*}({\bf r}\sigma, {\bf r\rq{}}\sigma\rq{})}.
\end{equation}
Introducing Bogoliubov transformation, which allows to express both normal and anomalous densities
in a form similar to the orbital expansion in conventional DFT,
one arrives at Kohn-Sham scheme for superfluid fermion systems, which
formally resembles the Bogoliubov-de Gennes equations.
Unfortunately they form set of integro-differential equations in coordinate space and their solution 
is extremely difficult in practice. 
This is a consequence of the nonlocality of the pairing potential $\Delta ({\bf r}\sigma, {\bf r\rq{}}\sigma\rq{})$.
It is possible however to formulate the problem using local pairing field \cite{Kur1999}. The justification for the so-called   
SLDA (Superfluid Local Density Approximation) has been developed in a series of papers 
(see Refs.~\cite{BY:2002fk,Bulgac:2002uq,BY:2003,Yu:2003,Bulgac:2007a, bulgac2013} )
and was shown to be very accurate for nuclei and cold atomic gases. 
The prescription involves the renormalization of the pairing coupling constant, which is a function of the momentum cutoff.
In the case of the spherical cutoff the analytic formula can be derived (spin indices are omitted for clarity):
\begin{eqnarray}
\Delta({\bf r}) &=& g_{eff} ({\bf r})\chi_{c}({\bf r}) \\
\frac{1}{g_{eff} ({\bf r})} &=& \frac{1}{g ({\bf r})} - \frac{m k_{c}({\bf r})}{2\pi^{2}\hbar^{2}}
\left ( 1 - \frac{ k_{F}({\bf r}) }{ 2k_{c}({\bf r}) } \ln \frac{ k_{c}({\bf r}) + k_{F}({\bf r})  }{ k_{c}({\bf r}) - k_{F}({\bf r}) }  \right )  ,
\end{eqnarray}
where anomalous density $\chi_{c}$ is defined within the truncated space and
\begin{eqnarray}
E_{c}+\mu &=& \frac{\hbar^{2}k_{c}^{2}({\bf r})}{2m({\bf r})} + \Gamma({\bf r}) \\
\mu &=& \frac{\hbar^{2}k_{F}^{2}({\bf r})}{2m({\bf r})} + \Gamma({\bf r}).
\end{eqnarray}
In the above formula $E_{c}$ defines the corresponding energy cutoff.

In a similar manner like in the conventional DFT, SLDA  can be extended to describe time dependent phenomena.
In this case the adiabatic approximation is applied, which neglects possible memory effects in the time evolution.
As a result the time-dependent Kohn-Sham equations have formal structure of the time dependent Bogliubov-de Gennes (TDBdG) or 
Hartree-Fock-Bogoliubov (TDHFB)
equations (spin indices are omitted)\footnote{In the nuclear physics community the term 'HFB' is more frequently used, although
BdG eqs. have formally the structure of HFB eqs. in coordinate representation.}:
 \begin{eqnarray} \label{tdslda}
i\hbar\frac{\partial}{\partial t} 
\left  ( \begin{array} {c}
  U_{\mu}({\bf r},t)\\  
  V_{\mu}({\bf r},t)\\ 
\end{array} \right ) =
\left ( \begin{array}{cc}
h({\bf r},t)&\Delta({\bf r},t)\\
\Delta^*({\bf r},t)&-h^* ({\bf r},t)
\end{array} \right )  
\left  ( \begin{array} {c}
  U_{\mu}({\bf r},t)\\
  V_{\mu}({\bf r},t)\\ 
\end{array} \right ),
\end{eqnarray}
where $h({\bf r},t) = -\frac{\hbar^{2}}{2m}\nabla^{2} + \Gamma({\bf r},t)$.

Using the above framework which is an extension of the DFT to the real-time
dynamics of Fermi superfluids it has been possible to describe a completely different physical system,
where pairing plays an important role - cold atomic gas in the so-called unitary regime.
Apart from being able to describe correctly known experimental 
facts, this approach leads also to new qualitative predictions including supercritical 
flow, quantum shock waves and domain walls, Higgs modes, vortex crossings, etc. (see Ref.~\cite{bulgac2013a, bulgac2013b, bulgac2013c, bulgac2013d}).

In the case of nuclear system the set of equations have to be solved for both protons and neutrons, which are coupled
through the potential $\Gamma({\bf r},t)$ depending on both neutron and proton densities.
For the case of Skyrme parametrization of the density functional the following local densities and currents are used as building blocks
(time variable is omitted):
\begin{itemize}
\item density: $\rho({\bf r})=\rho({\bf r},{\bf r'})|_{r=r'}$ ,
\item spin density: $\vec{s}({\bf r})=\vec{s}({\bf r},{\bf r'})|_{r=r'}$,
\item current: $\vec{j}({\bf r})=\frac{1}{2i}(\vec{\nabla}-\vec{\nabla}')\rho({\bf r},{\bf r'})|_{r=r'}$,
\item spin current (2nd rank tensor): ${\bf J}({\bf r})=\frac{1}{2i}(\vec{\nabla}-\vec{\nabla}')\otimes\vec{s}({\bf r},{\bf r'})|_{r=r'}$,
\item kinetic energy density: $\tau({\bf r})=\vec{\nabla}\cdot\vec{\nabla}'\rho({\bf r},{\bf r'})|_{r=r'}$,
\item spin kinetic energy density: $\vec{T}({\bf r})=\vec{\nabla}\cdot\vec{\nabla}'\vec{s}({\bf r},{\bf r'})|_{r=r'}$,
\end{itemize}
where both normal and anomalous densities are defined as:
\begin{eqnarray}
\rho({\bf r}\sigma, {\bf r'}\sigma')=
      \sum_{\mu}V_{\mu}^{*}({\bf r}\sigma)V_{\mu}({\bf r'}\sigma') ,\\
\chi({\bf r}\sigma, {\bf r'}\sigma')=
      \sum_{\mu}V_{\mu}^{*}({\bf r}\sigma)U_{\mu}({\bf r'}\sigma') ,
\end{eqnarray}
and the $U$ and $V$ components can be thought of as coefficients of the Bogoliubov transformation between single-particle ($\psi$)
and quasiparticle ($\alpha$) bases:
\begin{eqnarray}
\left  ( \begin{array} {c}
  \vec{\hat{\psi}}     \\
  \vec{\hat{\psi}}^{+}\\ 
\end{array} \right ) = {\cal B} 
\left  ( \begin{array} {c}
  \vec{\hat{\alpha}}     \\
  \vec{\hat{\alpha}}^{+}\\ 
\end{array} \right );  
 {\cal B} = 
 \left ( \begin{array}{cc}
U&V^{*}\\
V&U^* \\
\end{array} \right ) . 
\end{eqnarray}

The nuclear components of the potentials $\Gamma$ and $\Delta$ in the Skyrme parametrization read:
\begin{eqnarray}
&h({\bf r},t)&= -\vec{\nabla}\cdot \left  ( B({\bf r},t) + \vec{\sigma}\cdot\vec{C}({\bf r},t) \right )\vec{\nabla} 
+ U({\bf r},t) + \vec{U}_\sigma({\bf r},t)\cdot\vec{\sigma} \nonumber \\
&+& \frac{1}{2i} \left [\vec{W}({\bf r},t) \cdot (\vec{\nabla} \times \vec{\sigma} )+
                              \vec{\nabla} \cdot (\vec{\sigma} \times \vec{W}({\bf r},t)) \right]
+\frac{1}{i} \left ( \vec{\nabla}\cdot\vec{U}_\Delta({\bf r},t) 
+ \vec{U}_\Delta({\bf r},t)\cdot\vec{\nabla} \right ),
\end{eqnarray}
where
\begin{eqnarray}
B({\bf r},t) &=& \frac{\hbar^{2}}{2m} + C^{\tau}\rho \\
\vec{C}({\bf r},t) &=& C^{sT}\vec{s} \\
U({\bf r},t) &=& 2C^{\rho}\rho + 2C^{\Delta\rho}\nabla^{2}\rho + C^{\tau}\tau + C^{\nabla J}\vec{\nabla}\cdot\vec{J} + 
C^{\gamma}(\gamma+2)\rho^{\gamma+1} \\
\vec{W}({\bf r},t) &=& -C^{\nabla J} \vec{\nabla} \rho\\
\vec{U}_\sigma({\bf r},t) &=& 2C^{s}\vec{s} + 2C^{\Delta s }\nabla^{2}\vec{s} +  C^{sT}\vec{T} + 
C^{\nabla J}\vec{\nabla}\times\vec{j} \\
\vec{U}_\Delta({\bf r},t) &=& C^{j}\vec{j} + \frac{1}{2}C^{\nabla j}\vec{\nabla}\times\vec{s}
\end{eqnarray}
and pairing potential (spins are omitted):
\begin{equation}
\Delta({\bf r},t) = g_{eff}({\bf r},t) \chi({\bf r},t).
\end{equation}
Consequently the energy density functional reads
\begin{eqnarray}
E=\int d^{3} r {\cal H}({\bf r})
\end{eqnarray}
where
\begin{eqnarray}
{\cal H}({\bf r}) &=& C^{\rho}\rho^{2} + C^{s}\vec{s}\cdot\vec{s} + C^{\Delta\rho}\rho\nabla^{2}\rho + 
C^{\Delta s }\vec{s}\cdot\nabla^{2}\vec{s} + C^{\tau} (\rho\tau - \vec{j}\cdot\vec{j}) + C^{\nabla s}(\vec{\nabla}\cdot\vec{s})^{2} + \nonumber  \\
&+& C^{sT}(\vec{s}\cdot\vec{T} - {\bf J}^{2}) +
C^{\nabla J}( \rho\vec{\nabla}\cdot\vec{J} + \vec{s}\cdot(\vec{\nabla}\times\vec{j}) ) +
 C^{\gamma}\rho^{\gamma} - g_{eff}|\chi|^{2}
\end{eqnarray}
where
\begin{eqnarray}
J_{i} &=& \sum_{k,l}\epsilon_{ikl}{\bf J}_{kl} \\
{\bf J}^{2} &=& \sum_{k,l}{\bf J}_{kl}^{2}.
\end{eqnarray}
Apart from nuclear components the Coulomb term has to be added for protons.

One may express the equations (\ref{tdslda}) using the generalized density matrix which (in an arbitary basis) is defined as:
 \begin{eqnarray}
{\cal R}(t) =
\left ( \begin{array}{cc}
\rho &\chi \\
\chi^{+}&1-\rho^{*} 
\end{array} \right )  
\end{eqnarray}
and fulfills the equation of motion:
\begin{eqnarray}
i\hbar\frac{\partial}{\partial t} {\cal R}(t) = [ {\cal H}, {\cal R}],
\end{eqnarray}
where
 \begin{eqnarray}
{\cal H}(t) =
\left ( \begin{array}{cc}
h(t)             & \Delta(t)\\
\Delta^{+}(t)& -h^* (t)
\end{array} \right ).
\end{eqnarray}

The time evolution of the superfluid system governed by eqs. (\ref{tdslda}) exhibits fundamental differences 
as compared to the conventional TDDFT. In the latter
case one needs to evolve the number of orbitals equal to the number of particles forming the system
and consequently the complexity of the problem scales  with the number of particles.
In the TDSLDA however, all orbitals, that span Hilbert space defined by the Bogoliubov transformation, have to be evolved. 
This ensures that the total energy of the system is conserved during time evolution i.e. $\frac{d}{dt} E=0$.
If the space is truncated eg. by introducing a momentum or energy cutoffs at some initial time, then certain 
properties of the Bogoliubov transformation do not hold. Namely, the Bogoliubov transformation becomes noninvertible and
the closure relation is violated: ${\cal B}{\cal B}^{+} \neq 1$. As a consequence:
\begin{eqnarray}
\chi &\neq& -\chi^{T} \\
1 - \rho &\neq& UU^{+} ,
\end{eqnarray}
and the total energy is no longer conserved. However this problem becomes important only for a sufficently long time evolution,
and depends on both the strength of the pairing potential and the size of the subspace.
Weaker pairing admits longer evolution without violation of the energy conservation.
The total energy is still conserved to high accuracy if the evolution of the system is short enough.
The requirement for the truncated subspace to be large enough indicates that the complexity of TDSLDA increases rapidly and
does not scale with the number of evolved particles as in the case of standard TDDFT. If the equations (\ref{tdslda})
are solved in a box with certain discretization of spatial coordinates then it is the size of the box which determines the
complexity of the problem. Consequently the TDSLDA equations are usually $2-3$ orders of magnitude more computationally
demanding than standard TDDFT.

\section*{Numerical implementation}

The last decade has witnessed a proliferation of methods and techniques for numerically solving the Kohn-Sham equations (see Ref. \cite{beck:2000} and references therein).
Among these methods real-space and real-time methods applied to TDDFT turned out to be extremely efficient \cite{Yabana:1996}.
The mathematical structure of TDSLDA requires solving a system of coupled, complex, time-dependent nonlinear 
partial differential equations. In various applications of TDSLDA,
a spatial three-dimensional Cartesian grid in coordinate space with periodic boundary
conditions has been used, with derivatives evaluated in momentum (Fourier-transformed) space.
This method represents a flexible tool to describe large amplitude nuclear motion as it contains the coupling 
to the continuum and between single-particle and collective degrees of freedom.
The solutions are $U$ and $V$ components of the Bogoliubov transformation represented on a discrete three-dimensional spatial lattice. 
From the wave functions one can extract various observables using the usual quantum mechanical rules 
(velocity and pairing fields, density distributions, etc.).
 The time evolution is performed using the fifth-order predictor-corrector-modifier Adams-Bashforth-Milne (ABM) method, 
which provides a combination of high accuracy and numerical stability. The time step is usually chosen so the relative truncation error 
in the ABM method is between $10^{-7}$ and  $10^{-15}$.
The present computational abilities allow to consider boxes of sizes up to $50^{3} - 60^{3}$, which with the lattice constant of $1$ fm allow
to consider dynamics of arbitrarily heavy nuclei.
The time step of the order of $0.07-0.08\ \textrm{fm}/c$ 
ensures that the evolution for time intervals of the order of $10 000\ \textrm{fm}/c$ will keep the numerical accuracy within 
the accepted range. 

\section*{ Nuclear reactions within TDSLDA framework}

There are both bad and good news in regard to the application of TDDFT, and TDSLDA in particular, to nuclear reactions and in general
to any quantum scattering problem.
The good news is that TDSLDA is naturally suited to describe large amplitude collective motion of nuclear system
and offers a valuable insight into the dynamics described through the evolution of spatial density distribution.
This clear picture is missing in approaches operating within the energy representation. 
TDSLDA offers a computational framework which mimics closely
the way how the low energy nuclear scattering experiments are performed. Namely, it simulates the evolution of the system
in real time, where spatio-temporal coordinates of the collision/reaction can be easily extracted. It provides information 
about the energy distribution among various degrees of freedom, e.g. various types of nuclear deformations.

The bad news is related to the general difficulty of TDDFT to address questions concerning many-body wave functions.
Observables that require knowledge of the many-body wave function are not easy to extract. For example the state-to-state
transition probability: 
\begin{equation}
S_{i f}=\lim_{t\rightarrow\infty}\langle\Phi_{f}|\Phi (t)\rangle \mbox{, and} \lim_{t\rightarrow -\infty}|\Phi (t)\rangle = |\Phi_{i}\rangle
\end{equation}
is an important quantity, which however in the case of TDDFT requires a special procedure to compute\cite{Roh2004}.
Another difficult quantities are e.g. the momentum distribution  or the transitional densities which require more information 
than just the local densities which enter the expression for the energy functional \cite{Wilken:2007,Li:2011}.
In general, all more than one-body observables, including various types of conditional probabilities
are difficult to determine within the presented formalism.
Adiabatic approximation is another limitation
preventing us from the proper description of dissipation effects, except for the one-body dissipation processes.

\subsection*{Coulomb excitation and gamma absorption}

The simplest type of reaction which can be studied within the TDSLDA formalism is the Coulomb scattering of two nuclei.
It is the simplest case, as it does not involve nuclear interaction between colliding nuclei and can serve as a textbook
example of a nucleus being a subject of an external time dependent perturbations originating from electromagnetic field.
The Coulomb excitation can naturally lead to the excitation of various collective modes including giant dipole and giant quadrupole
resonances \cite{Baur:1986,Beene:1990,Ritman:1993,Schmidt:1993,Mueller:1994,Grunschloss:1999}. 
It is can also provide a tool for studies of multiphonon nuclear states \cite{Emling:1993,Emling:1994,Boretzky:1996,Aumann:1998,Bertulani:1999,Boretzky:2003}.

In order for a collision to lead to nonadiabatic nuclear processes, the external potential
have to vary in time fast enough, which implies that the collision has to occur at relativistic energies. 
The interaction time
needs to be relatively short, and  for an efficient excitation of nuclear modes of
frequency $\omega$, the collision time $\tau_{coll}=b/\gamma v$ has to
fulfill the condition that $\omega\tau_{coll} \simeq 1$.  Here $b$ is
the impact parameter, $v$ is the projectile velocity, and $\gamma =
(1-v^2/c^2)^{-1/2}$ is the Lorentz factor.
For example, in the collision process: $^{238}U {+}  ^{238}U$ at 700 MeV/n, studied in Ref. \cite{Stet2015} the
collision time was of the order of $10$ fm/c for impact parameters of the order of the nucleus diameter. 

In the Coulomb scattering the excitation process is governed by the electromagnetic field which has to be built into
the framework of the TDDFT. This can be achieved using the requirement of the gauge invariance of the energy density functional.
Namely, the coupling of the nuclear system to the electromagnetic field:
\begin{eqnarray}
\vec{E} &=& -\vec{\nabla}\phi - \frac{1}{c}\frac{\partial\vec{A}}{\partial t} , \\
\vec{B} &=& \vec{\nabla}\times\vec{A} \\
\end{eqnarray}
is realized through the following transformation:
\begin{eqnarray}
\vec{\nabla}\psi &\rightarrow& \vec{\nabla}_{A}\psi  =  \left (\vec{\nabla} - i\frac{e}{\hbar c}\vec{A}\right)\psi , \\
\vec{\nabla}\psi^{*} &\rightarrow& \vec{\nabla}_{-A}\psi^{*} =  \left(\vec{\nabla} + i \frac{e}{\hbar c} \vec{A}\right)\psi^{*} ,\\
i\hbar\frac{\partial}{\partial t}\psi  &\rightarrow& \left ( i\hbar\frac{\partial}{\partial t} - e\phi\right ) \psi ,
\end{eqnarray}
which implies that $\vec{\nabla}\psi\psi^{*} \rightarrow  \vec{\nabla}\psi\psi^{*}$.
Consequently 
it requires the following transformation of proton densities and currents (subscript $A$ denotes
the quantities in the presence of electromagnetic field):
\begin{itemize}
\item density: $\rho_{A}({\bf r})=\rho_{A}({\bf r})$,
\item spin density: $\vec{s}_{A}({\bf r})=\vec{s}({\bf r})$
\item current: $\vec{j}_{A}({\bf r})=\vec{j}({\bf r}) - \frac{1}{\hbar c}\vec{A}\rho({\bf r})$,
\item spin current (2nd rank tensor): ${\bf J}_{A}({\bf r})={\bf J}({\bf r}) - \frac{1}{\hbar c}\vec{A}\otimes\vec{s}({\bf r})$,
\item spin current (vector): $\vec{J}_A({\bf r})  = \vec{J}({\bf r}) -\frac{1}{\hbar c} \vec{A} \times \vec{s}({\bf r})$ ,
\item kinetic energy density: $\tau_{A}({\bf r})=
\left  (\vec{\nabla} - i\frac{1}{\hbar c}\vec{A}\right)\cdot\left (\vec{\nabla}' + i\frac{1}{\hbar c}\vec{A}\right)\rho({\bf r},{\bf r'})|_{r=r'} \\
 = \tau({\bf r}) -2\frac{1}{\hbar c}\vec{A}\cdot\vec{j}({\bf r}) +\frac{e^2}{\hbar^2 c^2}|\vec{A}|^2\rho({\bf r})   
 = \tau({\bf r}) -2\frac{1}{\hbar c}\vec{A}\cdot\vec{j}_A({\bf r}) -\frac{e^2}{\hbar^2 c^2}|\vec{A}|^2\rho({\bf r})$,
\item spin kinetic energy density: $\vec{T}_A({\bf r})=
\left (\vec{\nabla} - i\frac{1}{\hbar c}\vec{A}\right)\cdot\left (\vec{\nabla}' + i\frac{1}{\hbar c}\vec{A}\right)\vec{s}({\bf r},{\bf r'})|_{r=r'}\\
 =\vec{T}({\bf r}) -2\frac{1}{\hbar c}\vec{A}^T \cdot {\bf J}({\bf r}) +\frac{e^2}{\hbar^2 c^2}|\vec{A}|^2\vec{s}({\bf r}) 
 =\vec{T}({\bf r}) -2\frac{1}{\hbar c}\vec{A}^T \cdot {\bf J}_A({\bf r}) -\frac{e^2}{\hbar^2 c^2}|\vec{A}|^2\vec{s}({\bf r})$ .
\end{itemize}

As a result the proton single-particle hamiltonian acquires the form:
\begin{eqnarray}
h_{A}({\bf r},t)&=& -\vec{\nabla}_{A}\cdot \left  ( B({\bf r},t) + \vec{\sigma}\cdot\vec{C}({\bf r},t) \right )\vec{\nabla}_{A} 
+ \frac{1}{2i}\left ( \vec{W}({\bf r},t)\cdot( \vec{\nabla}_A\times\vec{\sigma} ) + 
\vec{\nabla}_A\cdot( \vec{\sigma}\times\vec{W}({\bf r},t) ) \right ) \nonumber \\
&+& U_{A}({\bf r},t) +\vec{U}_{\sigma}^{A}({\bf r},t)\cdot\vec{\sigma} 
+\frac{1}{i} \left ( \vec{\nabla}_{A}\cdot\vec{U}_{\Delta}^{A}({\bf r},t) + \vec{U}_{\Delta}^{A}({\bf r},t)\cdot\vec{\nabla}_{A} \right ) + e\phi ,
\end{eqnarray}
and
\begin{eqnarray}
&&U_{A}({\bf r},t) =  U({\bf r},t) - C^{\nabla J} \frac{1}{\hbar c}  \vec{\nabla}\cdot[\vec{A}\times\vec{s}] 
-C^{\tau}\left ( 2\frac{1}{\hbar c} \vec{A}\cdot \vec{j}+\frac{e^2}{\hbar^2 c^2}|\vec{A}|^2\rho \right ) , \\
&&\vec{U}_{\sigma}^{A}({\bf r},t) =
\vec{U}_\sigma({\bf r},t) - C^{\nabla J}\frac{1}{\hbar c}\vec{\nabla}\times [\vec{A}\rho ]
-C^{sT}\left ( 2\frac{1}{\hbar c} \vec{A}^T\cdot {\bf J}+\frac{e^2}{\hbar^2 c^2}|\vec{A}|^2\vec{s} \right ) , \\
&&\vec{U}_{\Delta}^{A}({\bf r},t) =   \vec{U}_{\Delta}({\bf r},t)  -C^{j} \frac{1}{\hbar c} \vec{A}\rho , \\
&&\vec{\nabla}_{A}\cdot \left  ( B({\bf r},t) + \vec{\sigma}\cdot\vec{C}({\bf r},t) \right )\vec{\nabla}_{A} =  
\left[ \vec{\nabla}_A \left  ( B({\bf r},t) + \vec{\sigma}\cdot\vec{C}({\bf r},t) \right )\right ]\cdot \vec{\nabla}_A + \nonumber \\
&&\left  ( B({\bf r},t) + \vec{\sigma}\cdot\vec{C}({\bf r},t) \right )
\left [ \Delta -i\frac{1}{\hbar c} \left ( \vec{A}\cdot \vec{\nabla}_A +\vec{\nabla}_A\cdot \vec{A} \right )  + \frac{e^2}{\hbar^2c^2}|\vec{A}|^2 \right ] ,
\end{eqnarray}
where spatial and time variables of quantities: $\rho, \vec{s}, \vec{j}, {\bf J}, \vec{A}, \phi$ have been omitted.

Apart from the coupling of proton charges and currents to the electromagnetic field, there is also a component which describes 
the interaction of magnetic field with
the nucleon spin: $\mu_{i}\vec{\sigma}\cdot\vec{B}$, where $\mu_{p}=5.5858 e\hbar/2m_{p}c$ and $\mu_{n}=-3.8263 e\hbar/2m_{p}c$
for protons and neutrons, respectively. This correction however is small and will be neglected.

One of the most straightforward applications of the TDDFT is the calculation of the nuclear photoabsorption cross section.
The process is an example of a perturbation of a nuclear system induced by a photon absorption, which can be described within the linear regime and although it 
does not require the whole machinery of TDDFT, it is not a trivial effect,
as it combines in a nutshell several challenging aspects of physics of an atomic nucleus \cite{harakeh}.
A perturbed nucleus exhibits large amplitude nonadiabatic motion and damping effects lead to a collective energy dissipation \cite{bertsch}. 
In the classical picture the isovector Giant Dipole Resonance (IGDR) is formed by two types of fluids representing neutrons and protons vibrating around 
a common center of mass.
In the Steinwedel-Jensen and Goldhaber-Teller models the mass dependence of the excitation energy reads: $A^{-1/3}$ and $A^{-1/6}$, respectively \cite{ring}. 
A reasonably good estimation of the IGDR vibrational frequency is $\omega \approx 80 MeV A^{-1/3}$ for spherical nuclei. 
For deformed nuclei, the IGDR reveals the splitting of characteristic frequencies which, roughly speaking, measures the aspect ratio of the nuclear shape. 
Due to the fact that IGDR is not an eigenstate of nuclear Hamiltonian, it is characterized by a spreading width which cannot be extracted from 
the hydrodynamical approach and has to be reproduced by including a reliable microscopic model of the atomic nucleus. 
The total width of IGDR possesses two components: one related to the coupling of the IGDR to more complex nuclear configurations
$\Gamma^{\downarrow}$, and the second associated with coupling to the continuum, e.g. related to emission lifetimes of particles (neutrons) $\Gamma^{\uparrow}$.
These two widths contribute to the total width of the IGDR, $\Gamma = \Gamma^{\downarrow} + \Gamma^{\uparrow}$, and their relative contributions vary depending on the mass number and the $N/Z$ ratio. 
The escape width is usually more important for light nuclei.
The physical mechanism related to $\Gamma^{\downarrow}$ may be quite complicated and depends on the energy. It involves coupling to low energy surface vibrations, Landau damping 
and collisional damping \cite{bertsch}.

In contemporary approaches, the description of the atomic nucleus is provided by DFT
and the IGDR has to be described within the same framework. The linear response formalism has been intensively used in the past due to the relative computational simplicity.
It originates from the small amplitude perturbation of the static DFT solution. 
This approach gives rise to the well known (Q)RPA equations that have been solved for a variety of functionals and forces, though only very recently for deformed systems \cite{terasaki,losa,peru,martini,stoitsov,avogadro}.
The application of full TDDFT formally incorporates the description of nonequilibrium phenomena and therefore covers both nonadiabatic and anharmonic
effects. In order to calculate the cross section for the photoabsorption cross section for deformed nuclei one needs to average results over
various orientations. It is usually a computationally demanding task especially in the case of triaxial nucleus like eg. $^{188}Os$
which need to be averaged at least over three different orientations.
The results obtained for various deformed systems: $^{172}Yb$, $^{188}Os$ and $^{238}U$ reveal a good agreement with experimentally extracted behaviour
of the cross section as a function of energy \cite{Stet2011}. The calculation of the width of the resonance requires however the proper treatment
of the dissipation effects and therefore cannot be reproduced.   

Another feature that one observes in a large amplitude nuclear motion studied within superfluid TDDFT (TDSLDA) is that the occupation
probabilities of proton and neutron quasiparticle states vary in time  considerably.
Namely, from the TDSLDA solutions one can extract the occupation probabilities for both proton and neutron quasiparticle states as follows:
\begin{equation}
n_k(t)=\sum_{\sigma=\uparrow,\downarrow}\int d^3r |V_k({\bf r},\sigma,t)|^2,
\end{equation}
where $k$ labels the proton and neutron quasiparticle wave functions respectively, which are solutions of the TDSLDA equations.
These occupation probabilities vary in time rather strongly and therefore the assumption which is usually made, to simplify
the calculations, that they are frozen to their ground state values is violated \cite{Stet2011}. 
 
The advantage of the TDDFT approach is that it can be used also beyond the linear regime when the perturbation 
of a nucleus is arbitrarily strong.
Such a case occurs in ultrarelativistic Coulomb collisions at impact parameters of the order of nuclear diameters. 
The question which can be easily provided by TDDFT is the amount of energy transferred into internal degrees of freedom as a result of
the collision. In the coulex reaction involving $^{238}U$ studied in Ref. \cite{Stet2015} it turned out that approximately half of the total energy has been transferred 
to internal degrees of freedom at impact parameters: $14-20$fm. The other half was responsible for translational motion of a target nucleus.
It is instructive to compare this fully microscopic result with the simple model based on the Goldhaber-Teller model (GT).
Within the model it is assumed that both protons and neutrons are represented by rigid 
density distributions which can oscillate harmonically against each other. 
Thus the target nucleus possess only two types of degrees of freedom: those related to
the center-of-mass (CM) motion and those describing the internal harmonic excitation of GDR.
The comparison between the average energy transferred to the internal motion
of the target nucleus  obtained within TDSLDA and also within the simplified Goldhaber-Teller 
model shows that significantly more energy is deposited by the projectile within
the TDSLDA. Namely, for impact parameters $14-20$ fm GT model predicts only $40-60$\%
of TDSLDA energy transferred to internal degrees of freedom. 
The Goldhaber-Teller model is equivalent to the linear regime, assuming that all isovector transition strength is concentrated in
two sharp lines, corresponding to an axially deformed target. 
An exact linear response approach would therefore severely
underestimate the amount of the internal energy deposited, one reason being 
the non-linearity of the response, naturally incorporated in TDSLDA.  
The other reason being the fact that the present microscopic framework
describing the target allows for many degrees of freedom, apart
from pure dipole oscillations, to be excited.  At the same time, the CM target energy  
alone is approximately the same as obtained in a simplified point particles
Coulomb recoil model of both the target and projectile.

After collision the excited nucleus will subsequently emit radiation and neutrons.
Part of the radiation emitted right after collision can be described within the TDDFT.
Namely, having the proton densities and currents extracted from TDSLDA
\begin{eqnarray}
 \rho({\bf r},t) &=& \int_{-\infty}^{\infty} \frac{d\omega}{2\pi} \rho({\bf r},\omega) \exp (-i\omega t) , \\
 \vec{j}({\bf r},t) &=& \frac{1}{2\pi}\int_{-\infty}^{\infty} d\omega \vec{j}({\bf r},\omega) \exp (-i\omega t) ,
\end{eqnarray}
one may evaluate the frequency distribution of emitted radiation:
\begin{eqnarray}
\frac{dE}{d\omega}   =   \frac{4 e^2}{c} \sum_{l,m} |\vec{b}_{lm}(k,\omega)|^{2} 
\end{eqnarray}
and the radiated power
\begin{eqnarray} 
P(t + r/c) &=&  \frac{e^{2}}{\pi c } \sum_{l,m} \left | \int_{-\infty}^{\infty} \vec{b}_{lm}(k,\omega) \exp (-i \omega t ) d\omega \right |^{2}
\end{eqnarray}
where
\begin{eqnarray}
\vec{b}_{lm}(k,t)      &=& \int d^{3}r\vec{b}({\bf r},t)j_{l}(kr)Y_{lm}^{*}(\hat{r}) \\
\vec{b}_{lm}(k,\omega) &=& \int_{-\infty}^{\infty} \vec{b}_{lm}(k,t)\exp(i\omega t) dt
\end{eqnarray}
describe the corresponding multipole components of radiation.
In the calculations presented in Ref. \cite{Stet2015} the nuclear evolution has been followed
during approximately $2500$ fm/c after collision and two components
of the electromagnetic radiation can be distinguished. The one  which arises from the
CM acceleration as a result of collision (Bremsstrahlung), and
takes part only during the relatively short time interval
$\tau_{coll}={b}/{v\gamma}$. This contribution is of the order of $0.1\%$ of the total radiation emitted within the first $2500 $ fm/c.
The radiation emitted from the internal motion has much longer time scale.  It was concluded that 
the main part of radiation coming from the target nucleus is due
to the excitation of IGDR. The smaller fraction is related to the Giant Quadrupole Resonance (GQR) and also can be attributed to the low lying mode (pygmy resonance).
One needs to remember, however, that the amount of energy emitted during this time interval is of the order of $1$\% of
the total absorbed energy during the collision.
 It turned out that although the intensity of
radiation decreases with increasing impact parameter, the ratio between
the intensities, due to the internal modes with that of the CM motion,
remains fairly constant.

The evolution of the total dipole moment of the target nucleus can be easily extracted from TDSLDA. It exhibits damping which
is a consequence of one-body dissipation processes, leading to the transfer of the collective energy to single particle degrees of freedom.
This dissipation process can be studied in TDDFT, even within the adiabatic approximation. It provides however only a small fraction
of the total width of the collective mode.
Assuming that the energy of the dipole mode is proportional to the square of its amplitude one finds out that the
damping rate is almost perfectly described by the exponential decay law: $E_{coll}(t)\propto\exp{(-t/\tau)}$ with $\tau\approx 500$ fm/c and
it does not depend on the nucleus orientation during the collision \cite{Stet2015}.

\subsection*{Nuclear reactions, fusion and fission}

The most interesting processes, although still to large extent unexplored, are related to nuclear collisions and in particular to nuclear fission and fussion. 
Indeed the role of pairing correlations is regarded as the key ingredient, which allows to properly describe an induced nuclear fission and therefore
the superfluid extension of TDDFT seems to be the natural candidate for a microscopic theory of induced fission.
Fully microscopic description of nuclear fission, which is a long standing goal, is a research topic more 
than seven decades old, a problem of great practical and fundamental interest. Its complexity due to
the large number of strongly coupled degrees of freedom made this problem computationally challenging. 
It has been pointed out in the
papers by Meitner and Frisch and Bohr and Wheeler in 1939 \cite{Me1939,Bo1939} that nuclear fission can be 
regarded as the evolution of the nuclear shape leading eventually to splitting into two or more fragments, although
its dynamics is still not well established. The process of transforming the compound nucleus into fragments is not
well understood and usually the simplest adiabatic approximation is applied. The adiabatic approximation is however
a questionable assumption around the scission point where the collective motion of the nucleus speeds up.
Moreover the currently applied phenomenological methods introduce several parameters whose values are determined from adjustments to various observables. In most of these approaches the
nuclear shape evolves on the nuclear potential energy landscape, being a subject
to both conservative and dissipative forces until the scission point is reached. The location of this point is to certain extent arbitrary
and have rather obscure meaning in quantum mechanics.
At this point various observables including the kinetic energy of fragments and their masses are extracted.
As a consequence phenomenological approaches of this type have rather limited predictive power. 
The superfluid TDDFT, and TDSLDA in particular, offers the possibility to describe the fission process around the scission point without 
such assumptions, in particular the one
concerning adiabaticity, and thus can provide the quantitative predictions for the kinetic energy distributions of fragments.

It is important to emphasize  several aspects of superfluid TDDFT which need to be considered during the studies of nuclear reactions:
\begin{itemize}
\item The preparation of the initial condition needs to take into account the property that the $U$ components of the Bogoliubov transformation
are not localized within a nucleus and extend over the whole space. Therefore the preparation of the initial configuration is more complicated, 
as in the static solution the properties of $U$ components depend on the mutual arrangement of nuclei. Moreover one cannot simply apply
the Galilean boost to one nucleus without affecting the other through the change of the distribution of $U_{\mu}({\bf r})$.
The strategy which can be applied in this case involves separation of two fragments by a nonpenetrable potential wall, which is subsequently
removed during the evolution. This ensures that two nuclei are completely disentangled at the initial time.
\item The quantities like the cross section, which can be extracted from the calculations require an additional degree of freedom
to be averaged over. Namely, various relative phases of the pairing fields of two colliding nuclei need to be considered. It is {\em a priori} not
known whether this effect will have a significant impact on a collision process, but clearly certain observables, e.g. the particle transfer rate,
may turn out to be seriously affected.
\item The description of the particle transfer and the emission of more than one nucleon would be an interesting test for the quality of the 
nuclear energ density functional.
 It is well known for example that an analogue of such process in atomic physics, namely, the multiple ionization is hard to decribe within TDDFT \cite{Pet1999}.
\end{itemize}

\section*{Conclusions}

Nuclear reactions and fission in particular
 play an important role in applications to energy production,  astrophysics, etc. and in recent years the nuclear fission research has undergone a renaissance worldwide. Particularly important for applications are the properties of prompt fission neutrons and gamma rays, which are emitted before the weak decays of the fission fragments toward stability. Due to the complexity of the nuclear many body problem, the computationally realistic description of either nuclear reactions or nuclear structure properties required methods based on different assumptions which allowed to simplify the problem and made it tractable. Therefore the unified description of nuclei, both their static and dynamic properties is dramatically called for, as otherwise one would not be able to gain a deep physical insight into nuclear processes. 
The superfluid TDDFT, e.g. in the framework of TDSLDA is a perfect candidate to provide a fully microscopic description of nuclear fission and low energy nuclear reactions.
With increasing computational abilities it is likely  to become soon a standard tool in the field, having the advantage of being a fully microscopic theory and treating
all nucleonic degrees of freedom on the same footing.
In the next years we will certainly witness the growing importance of methods based on
time-dependent DFT, probing its advantages and limitations in the application to large amplitude nuclear dynamics - one of the greatest unsolved nuclear many-body 
problems.

\subsection*{Acknowledgments} 
I would like to thank my collaborators: Aurel Bulgac, Carlos Bertulani, Michael M. Forbes, Kenneth J. Roche, Ionel Stetcu, and Gabriel Wlaz{\l}owski.
I acknowledge support of Polish National Science Centre (NCN) Grant, decision 
no. DEC-2013/08/A/ST3/00708. This work has been also partly supported by the ERANET NuPNET
grant SARFEN of the Polish National Centre for Research and Development (NCBiR),
by Polish National Science Centre (NCN) Grant under Contract no. UMO-
2012/07/B/ST2/03907,  and U.S. Department of Energy (DOE) Grant No. DE-FG02-97ER41014.

\end{document}